\documentclass[aps,pra,twocolumn,superscriptaddress,nopacs,letterpaper]{revtex4}

\usepackage{amsmath}
\usepackage{color,graphicx}
\usepackage[pdfstartview=FitH,colorlinks=true,linkcolor=blue,citecolor=blue,urlcolor=blue]{hyperref}

\begin{document}

\title{Coherent Oscillations in Small Fermi Polaron Systems}

\author{Marek Tylutki}
\email{marek.tylutki@aalto.fi}
\affiliation{INO-CNR BEC Center and Dipartimento di Fisica, Universit\`a di Trento, 38123 Povo, Italy}
\affiliation{COMP Centre of Excellence, Department of Applied Physics, Aalto University, 00076 Aalto, Finland}

\author{G. E. Astrakharchik}
\affiliation{Departament de F\'{\i}sica, Universitat Polit\`{e}cnica de Catalunya, 08034 Barcelona, Spain}

\author{Alessio Recati}
\affiliation{INO-CNR BEC Center and Dipartimento di Fisica, Universit\`a di Trento, I-38123 Povo, Italy}
\affiliation{Ludwig-Maximilians-Universit\"{a}t M\"{u}nchen, 80333 M\"{u}nchen, Germany}

\date{\today}

\begin{abstract}
We study the ground state and excitations of a one-dimensional trapped polarized Fermi gas interacting with a single impurity. First, we study the tunnelling dynamics of the impurity through a potential barrier, such as one effectively created by a double-well trap. To this end, we perform an exact diagonalization of the full few-body Hamiltonian and analyze the results in a Local Density Approximation. Off-diagonal and one-particle correlation matrices are studied and are shown to be useful for discerning between different symmetries of the states. Second, we consider a radio-frequency (RF) spectroscopy of our system and the resulting spectral function. These calculations can motivate future experiments, which can provide a further insight into the physics of a Fermi polaron.
\end{abstract}

\maketitle

\section{Introduction}
\begin{figure}[!t]
\includegraphics[width=.7\columnwidth]{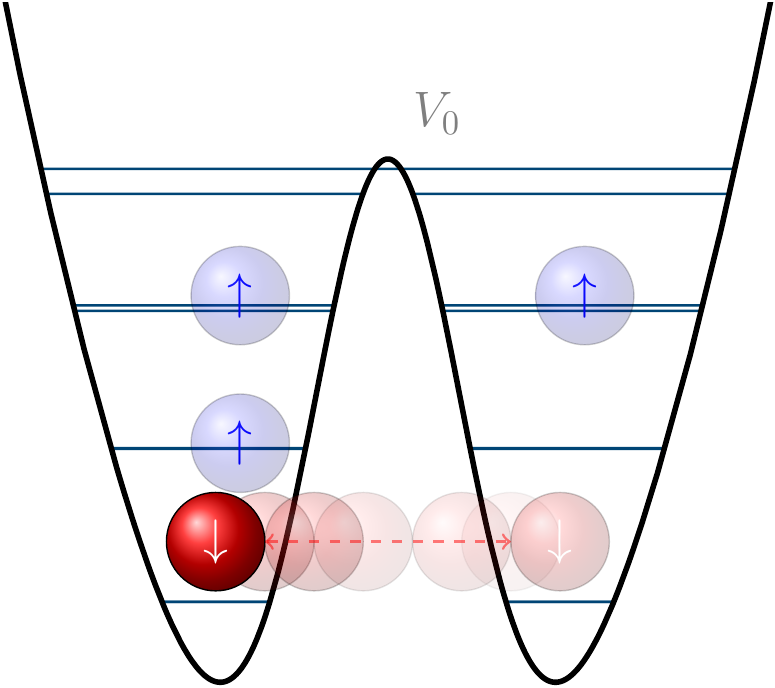}\\
\caption{(color online) Schematic representation of the system. 
A single impurity is oscillating in a double well potential, Eq.~(\ref{eq.dwpot}), but the oscillations are affected by the presence of a background gas of single-species fermions. The pair-wise quasi-degeneracy of the low lying single particle energy levels, where each pair contains states with an odd and an even wave function, is responsible for the {\it ``shell''} structure of the resulting spectrum.}
\label{fig.scheme}
\end{figure}

Interacting ultracold Fermi gases in continuum and trapped in optical lattices provide clean setups for experimental realizations of models essential for our understanding of the more complex condensed-matter systems. Since the interactions between fermions, combined with the Pauli exclusion principle, can lead to very complicated physics, it has always been instructive to simplify the system as much as possible and to consider a mostly non-interacting Fermi gas, where only one particle (an {\it impurity}) can interact with other atoms. It is even more instructive to study interactions of such an impurity with only a few atoms of the non-interacting species. What is more, a recent progress in confining ultracold gases and in detection techniques has allowed for trapping and studying just a few atoms with an unprecedented control. A single impurity atom interacting with a few identical fermions in one dimension (1D), schematically represented in Fig.~\ref{fig.scheme}, was experimentally realized, as described in Refs.~\cite{Wenz2013,Murmann2015}. The equation of state was measured starting with only one atom in each component and then increasing the number of majority atoms one by one. It was found that in one-dimensional geometry five atoms were already sufficient to find a good agreement between the measured polaron energy shift and the exact result for a single impurity interacting with a homogeneous Fermi gas by McGuire~\cite{McGuire1965,McGuire1966}. Such a quick arrival to the thermodynamic limit suggests that in a one-dimensional geometry a few fermions are already many. These experiments allow for studying the mesoscopic counterpart of the so-called Fermi polaron problem. The latter received much attention in the past decade thanks to a joint experimental and theoretical effort (see, e.g.,~\cite{PolaronReviewTeo} and references therein). Many properties such as the energy, effective mass, lifetime as well as some coherence properties of the polaron have been experimentally addressed. Evidence of a stable polaron quasi-particle on the repulsive branch has also been reported \cite{RudiTeo,Scazza2016}.
As far as the dynamics is concerned, the information about the dressed particle is carried by the relative motion of the impurity and the Fermi bath. For instance, the effective mass can be extracted by measuring out-of-phase modes~\cite{SalomonEffMass}. Yet, in the presence of a three-dimensional large bath, the relative dynamics consists of many strongly damped modes, and the dynamics of dressed impurity is often dominated by decaying processes~\cite{Bruun2008, SalomonEffMass}.

On the other hand, in a one-dimensional small bath, it is possible to have long living modes. Moreover, for a small number of particles, exact results for both static and dynamic properties can be obtained by exact diagonalization (ED). Such systems were already studied theoretically to some extent, beginning with an analytical solution for two harmonically trapped atoms by Busch {\it et al.}~\cite{Busch1998}. The limit of strong repulsion is very peculiar in 1D, as it effectively plays the role of the Pauli exclusion principle and leads to fermionization; it was investigated for several particles in Refs.~\cite{Garcia-March2014,Garcia-March2014a,Campbell2014}. In Ref.~\cite{Deuretzbacher2014} such a two-component Fermi system was studied with a mapping to an effective spin chain, and in Ref.~\cite{Volosniev2014} by constructing an energy functional. A numerical method for extracting the information about excitations corresponding to the relative motion of the particles in a harmonically trapped Fermi gas was developed recently~\cite{Pecak2017}. However, it is important to understand how the interaction with the background gas affects tunnelling properties of a polaron through a thin barrier, such as present in a double-well trap. Such traps were investigated for both bosons~\cite{Garcia-March2015,Dobrzyniecki2016} and fermions~\cite{Murmann2015,Sowinski2016}.

In the present work, we study the motion of an impurity coupled to a few identical Fermi atoms in 
a one-dimensional trap. In particular, we address the question of the frequency shift of the dipole mode of the 
impurity in a harmonic trap as well as the renormalization of the tunnelling frequency in the case 
of a double-well potential. We anticipate that also in the latter case an 
inhomogeneous extension~\cite{Astrakharchik2013} of the McGuire expression gives very reasonable 
results.

The paper is organized as follows. After defining our system in Sec.~\ref{sec.system}, we start with the description of the polaron system in the Local Density Approximation (LDA) with McGuire formula in Sec.~\ref{sec.results} and then compare it with the results obtained with ED. We study the dynamics of the oscillations in harmonic and double-well potentials in Sec.~\ref{sec.oscill}, and compare it with system's eigenenergies and with a sum-rule approach. Finally in Sec.~\ref{sec.rabi} the spectral function for the impurity spectroscopy is determined using the {\it Fermi's Golden Rule}.

\section{The system}
\label{sec.system}
We consider $N_\uparrow$ Fermi atoms interacting with one impurity atom denoted by the subscript 
$\downarrow$, whose dynamics is constrained to 1D and in presence of an external potential. 
The trapping potential can be either a standard harmonic confinement or a double-well potential.  
We model it as
\begin{equation}
V(x) = \frac{m \omega_0^2}{2} x^2 + V_0 e^{-m \omega_0 x^2 / 2 \hbar} + {C} ~,
\label{eq.dwpot}
\end{equation}
with the value of the constant offset $C$ such that the minimal value of the potential is equal to 
zero; the simple harmonic potential case corresponds to  $V_0=C=0$. 
Our system is schematically represented in Fig.~\ref{fig.scheme}. The many-body Hamiltonian can be written as
\begin{align}\nonumber
\mathcal{H} = \int_{-\infty}^\infty dx\, 
[ \sum_{\sigma=\uparrow,\downarrow}\psi_\sigma^\dag(x) h_0 \psi_\sigma(x) \\
+ g_{1D} \psi_\uparrow^\dag(x) \psi_\downarrow^\dag(x) \psi_\downarrow(x) \psi_\uparrow(x)] ~,
\label{eq.hamilt}
\end{align}
 where $\psi_\uparrow(x)$ is the field of the background polarized Fermi atoms and 
$\psi_\downarrow(x)$ is the field of the impurity. 
The field operators for the Fermi gas obey usual anticommutation relations: 
$\{ \psi_\uparrow(x), \psi_\uparrow^\dag(x^\prime) \} = \delta(x - x^\prime)$ 
and $\{ \psi_\uparrow(x), \psi_\uparrow(x^\prime) \} = 0$, 
whereas the statistics of the impurity does not matter. The operator $h_0= -\frac12 \hbar^2 \nabla^2 / m + V(x)$ stands for the single-particle Hamiltonian, while  $g_{1D} = -2 \hbar^2/ (m a_{1D})$ is the coupling constant in a homogeneous system in terms of the one-dimensional $s$-wave scattering length $a_{1D}$. When the dynamics is confined to 1D by a tight transverse harmonic confinement, the effective scattering length can be expressed in terms of the 3D scattering length, $a_{3D}$, and the oscillator length of the tight confinement, $\xi_\perp$, i.e. $a_{1D} \approx - \xi_\perp^2\, (1 - 1.46\, a_{3D} / \xi_\perp) / (2 a_{3D})$ as in Ref.~\cite{Olshanii1998}. In the following we use
\begin{equation}
\xi_0 = \sqrt{\hbar / (m \omega_0)} ~, \quad \varepsilon_0 = \hbar \omega_0 ~, \quad g_0 = \hbar \omega_0 \xi_0
\label{eq.units}
\end{equation}
as the units of length, energy and the coupling constant, respectively.

\section{Results}
\label{sec.results}
\subsection{Non-uniform McGuire formula}
\label{subsec.mcguire}
In order calculate the polaron energy, i.e. the shift of the energy of the system due to the interaction between the impurity and the polarized gas, we first adopt an approach based on the McGuire expression~\cite{McGuire1965,McGuire1966,Astrakharchik2013}, where the energy shift of the impurity in a uniform system is given by
\begin{equation}
\frac{\Delta E(N)}{E_F} = \frac{\gamma}{\pi^2} \left[ 1 - \frac{\gamma}{4} + \left( \frac{\gamma}{2 \pi} + \frac{2 \pi}{\gamma} \right) \arctan\frac{\gamma}{2 \pi} \right] ~.
\label{eq.mcguire}
\end{equation}
Here $E_F = \hbar^2 \pi^2 n^2/ (2 m)$ is the Fermi energy of the uniform non-interacting gas and $n$ its density. The ratio between the interaction and the kinetic energy, a.k.a. the Lieb-Liniger parameter
\begin{equation}
\gamma = \frac{\pi m g_{1D}}{\hbar^2 k_F} ~,
\label{eq.lieblinigergamma}
\end{equation}
determines the strength of interactions in a homogeneous system and reads as $\gamma = -2 / (n a_{1D})$ in terms of the one-dimensional $s$-wave scattering length $a_{1D}$. We generalize formula~(\ref{eq.mcguire}) to a non-uniform system by substituting the corresponding Fermi wave vector, $k_F$, in Eq.~(\ref{eq.lieblinigergamma}) according to
\begin{equation}
\mu_{\uparrow}(N_\uparrow) = \frac{\hbar^2}{2m}\, k_F^2,
\end{equation}
where $\mu_{\uparrow}$ is the chemical potential of the majority component (i.e. a polarized Fermi gas) in the external trap (either a harmonic oscillator or a double well) in the absence of the impurity. For a harmonic trap this expression takes a particularly simple form, $\mu_\uparrow = E(N_\uparrow) - E(N_\uparrow - 1) = N_\uparrow \hbar \omega_0$. 

\begin{figure}[!h]
\includegraphics[width=.9\columnwidth]{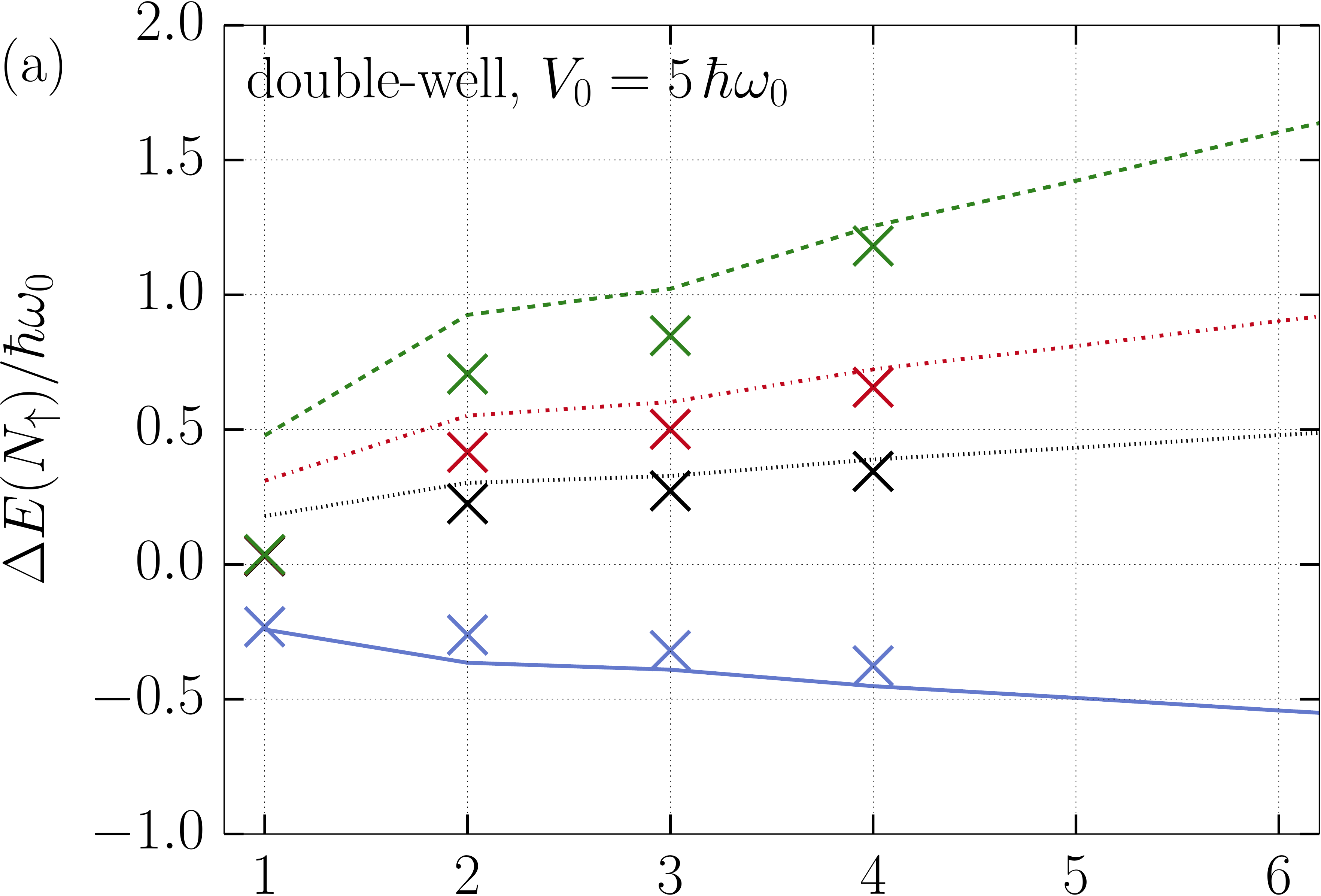}\\
\includegraphics[width=.9\columnwidth]{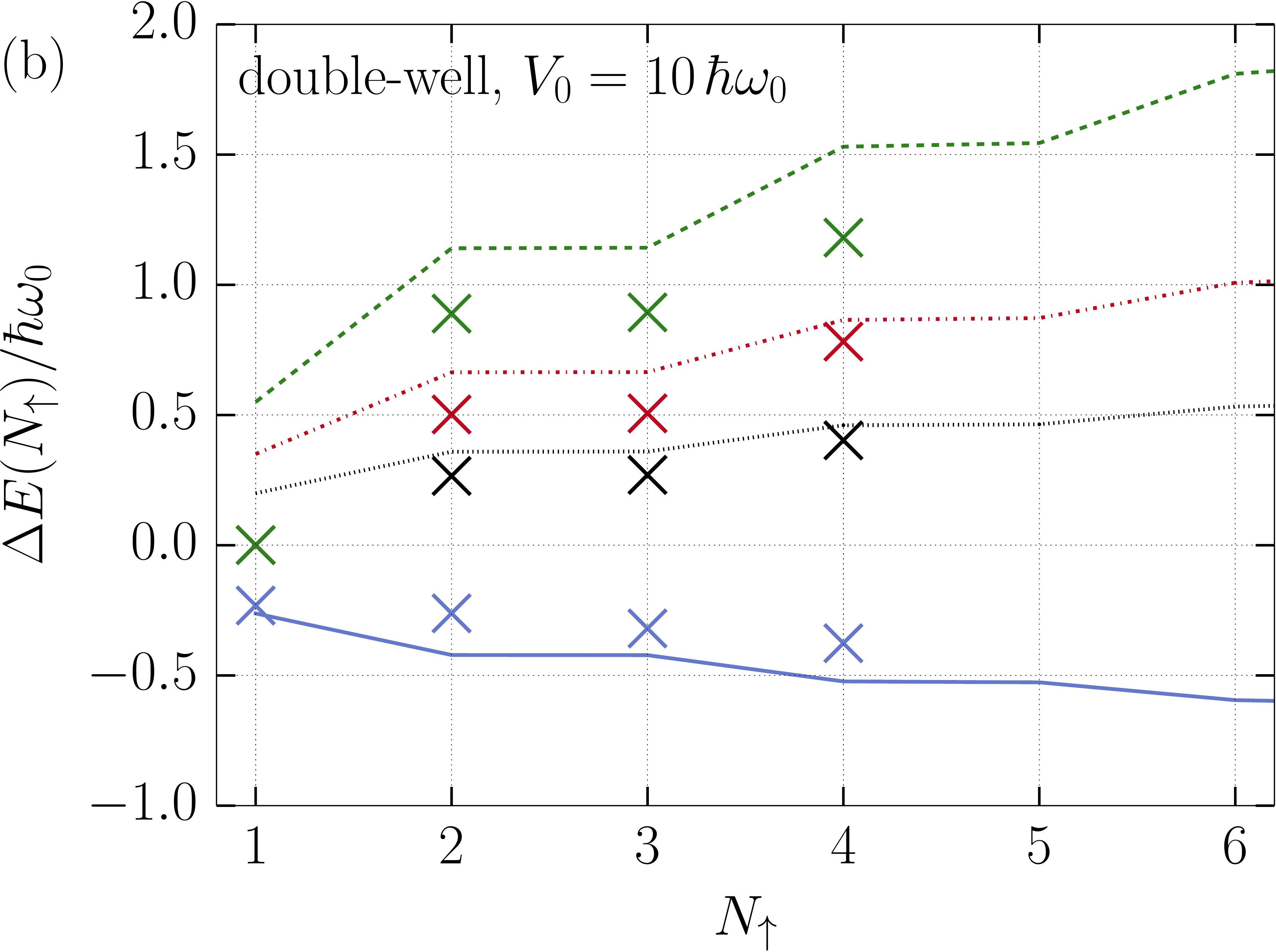}\\
\caption{(color online) Splitting of the first two energy levels in a double-well potential. Solid lines show a result obtained with McGuire formula adapted to a inhomogeneous geometry~(\ref{eq.mcguire}) and the crosses correspond to exact energies obtained by exact diagonalization. For legend see Fig.~\ref{fig.deltaEho}}
\label{fig.deltaEdw}
\end{figure}

The resulting polaron energy as a function of the number of majority atoms and for various strengths of the coupling constant $g_{1D}$ is shown in Figs.~\ref{fig.deltaEho} and~\ref{fig.deltaEdw} for a harmonic trap and a double-well potential respectively. The predictions of Eq.~(\ref{eq.mcguire}), shown with a dashed line, follow closely the results of ED (see the next Section), which are marked with crosses. 
\begin{figure}[!h]
\includegraphics[width=.9\columnwidth]{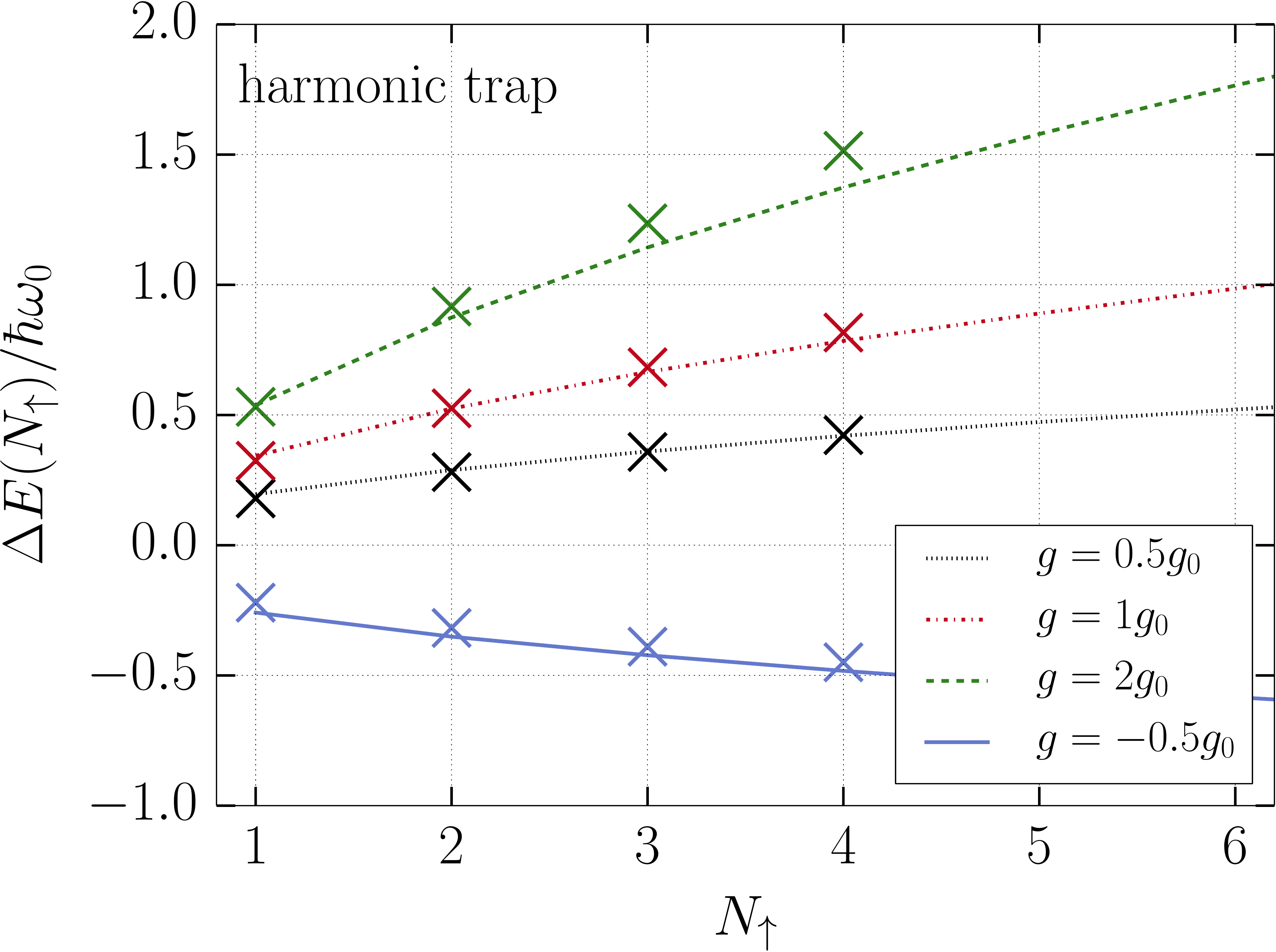}\\
\label{fig.deltaEho}
\caption{(color online) Polaron energy in a harmonic potential. Dashed lines show results obtained with McGuire formula adapted to a inhomogeneous geometry~(\ref{eq.mcguire}) and the crosses show the exact energies as obtained by exact diagonalization. For comparison, see Ref.~\cite{Astrakharchik2013}}
\label{fig.deltaEho}
\end{figure}

The corrected McGuire expression gives a very accurate result for a few body system in the presence of a harmonic confinement, as noted in~\cite{Astrakharchik2013}. Here, we find a good agreement in presence of a double-well potential (except for the case of $N_\uparrow = 1$ particles with repulsive interaction), which suggests that the generalization of the McGuire formula can be used for trapped impurities. In particular, for a double-well potential the energy shift of an impurity clearly shows formation of ``steps'' as the number of particles is increased one by one. This is due to almost degenerate structure of the low-lying energy levels. Because of the presence of the barrier and the symmetry of the potential under parity transformations ($x \to -x$), there is little difference between a state whose wave function has an extra node in the centre and a state whose wave function has one node less and is only exponentially suppressed by the barrier. This effect can be interpreted as a {\it ``shell''} structure with two fermions occupying each shell. The higher the barrier between the wells, the smaller the level splitting, because the wells are more independent. As a consequence, the Fermi energy grows with adding subsequent spin-$\uparrow$ atoms only after filling both levels in every pair. The discrepancy between the generalized McGuire expression and the exact result grows with the height of the potential barrier, because the energy levels are becoming pairwise {\it quasi}-degenerate, and these pairs are further apart from each other.

\subsection{Exact diagonalization}
\label{subsec.ed}
\begin{figure}[!h]
\includegraphics[width = .49\textwidth]{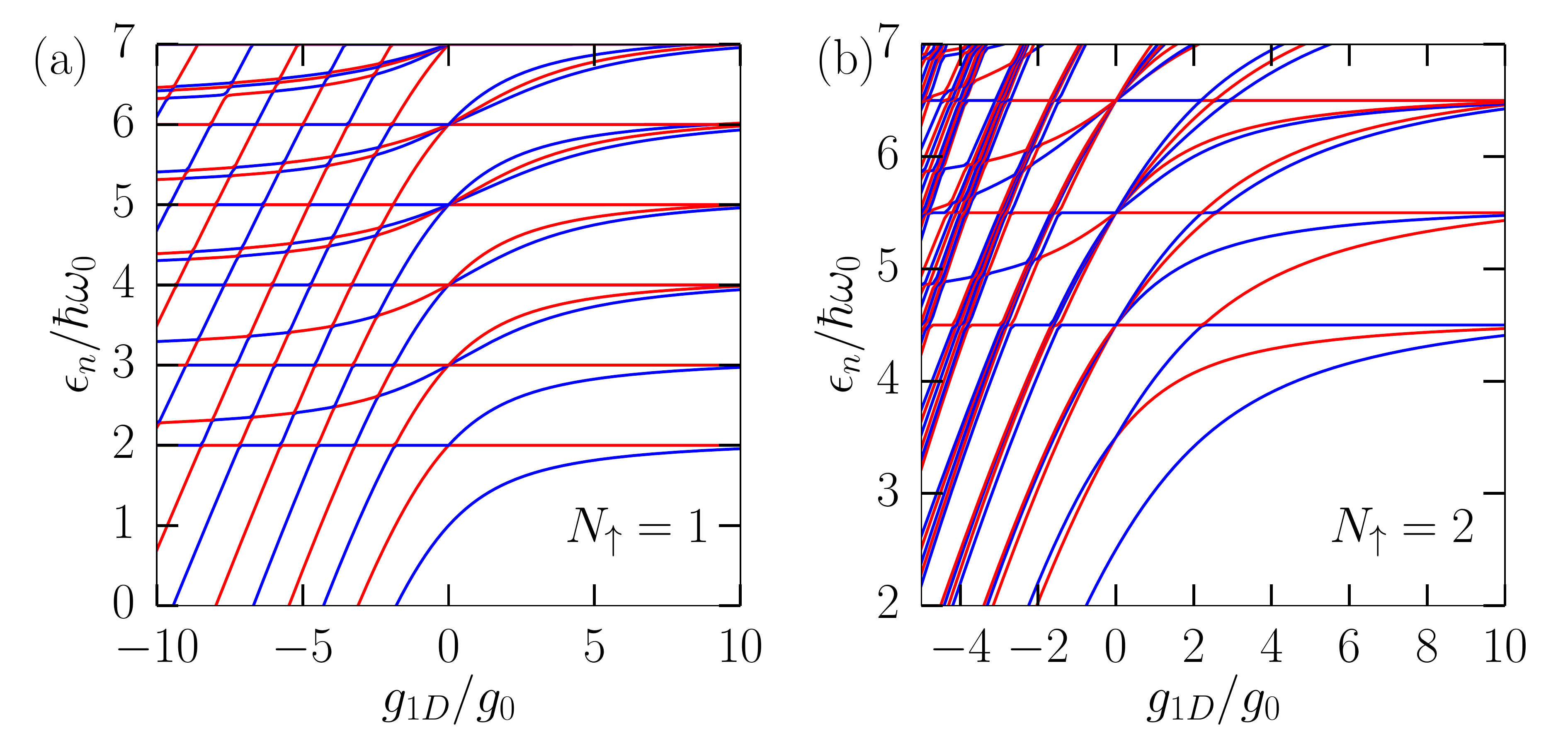}\\
\includegraphics[width = .49\textwidth]{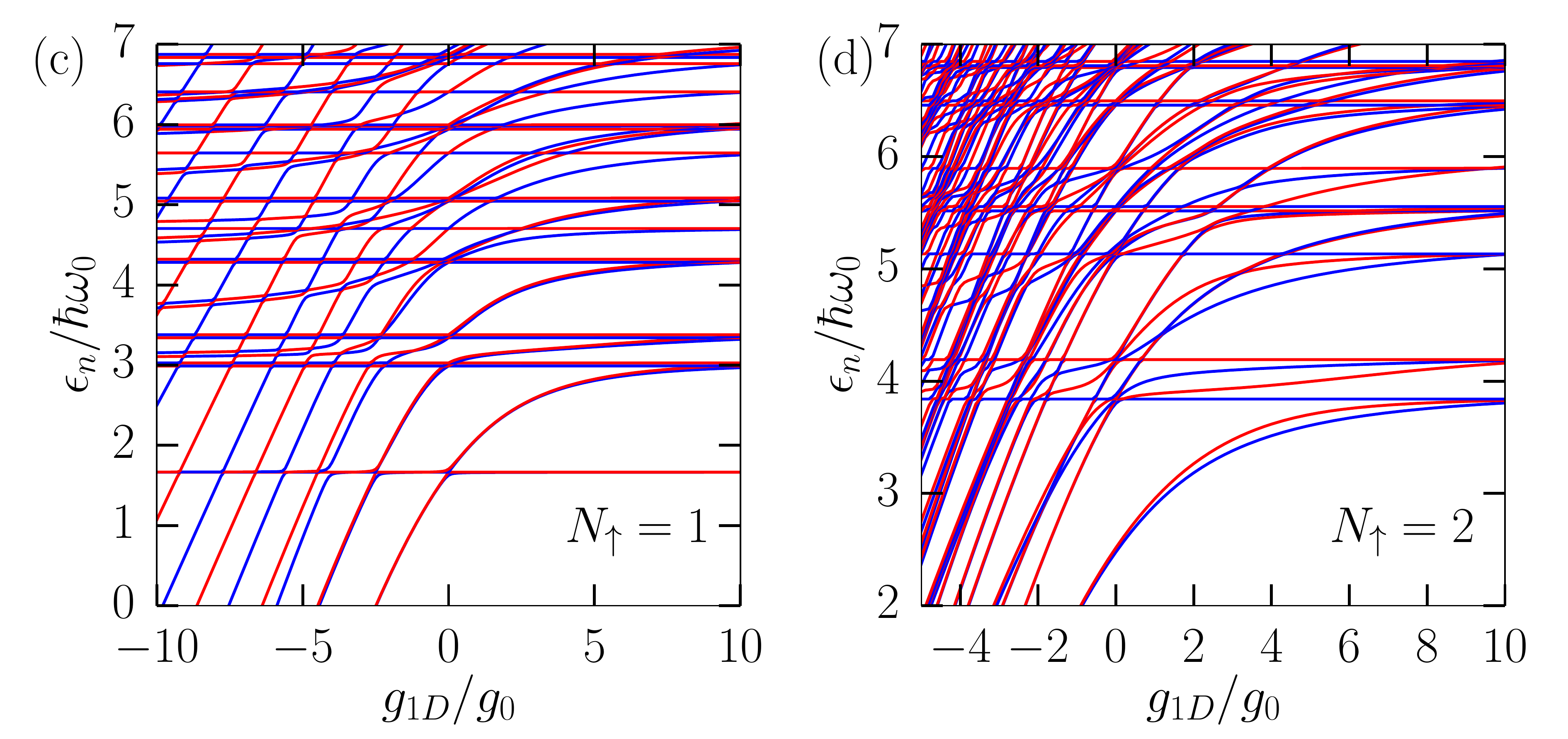}\\
\caption{(color online) (a) Spectrum of the $N_\uparrow = 1$ and (b) $N_\uparrow = 2$ problem in the h.o. potential; (c) Spectrum of the $N_\uparrow = 1$ and (d) $N_\uparrow = 2$ problem in the double-well potential for $V_0 = 5 \hbar \omega_0$. The lines are alternately coloured with blue and red for clarity. We can see, that the spectra for a double-well are pairwise almost degenerate (i.e. blue and red lines overlap) which reflects the structure of the single-particle spectrum.}
\label{fig.spectra}
\end{figure}

We study the system with the exact diagonalization, which is very useful for out-of-equilibrium 
phenomena~\cite{Sowinski2013,Dobrzyniecki2016,Pecak2016,Grass2015,Levinsen2015,Sowinski2016}. 
The field operators can be decomposed as: 
$\psi_\sigma(x) = \sum_n a_{\sigma, n} \varphi_n(x)$, with $\sigma = \uparrow, \downarrow$ and $\varphi_n$'s forming a complete eigenbasis of $h_0$, $h_0 \varphi_n = \epsilon_n \varphi_n$. The operators $a_{\uparrow n}$ and 
$a_{\downarrow n}$ are annihilation operators of $\uparrow$ and $\downarrow$ fermions in the single-particle state $n$, respectively. 
The Hamiltonian then reads
\begin{equation}
\label{HED}
\mathcal{H} = \sum_{i, \sigma} \epsilon_i a_{\sigma i}^\dag a_{\sigma i} + \sum_{ijkl} J_{ijkl}\, a_{\uparrow i}^\dag a_{\downarrow j}^\dag a_{\downarrow k} a_{\uparrow l}
\end{equation}
with $J_{ijkl} = g_{1D}\, \int_{-\infty}^\infty dx\, \varphi_i^*(x) \varphi_j^*(x) \varphi_k(x) \varphi_l(x)$. An example of energy spectra is shown in Fig.~\ref{fig.spectra} for harmonic oscillator and  double-well potentials.
The horizontal lines correspond to the fully antisymmetric states, which are therefore insensitive to interaction. For large positive values of the interaction strength, the impurity starts behaving like an additional fermion of the majority component, and the spectrum looks like that of a non-interacting gas of $N_\uparrow + 1$ fermions. This is known as a {\it fermionization limit} in analogy to the corresponding case of strongly repulsive bosons.

Before studying the dynamics, we determine the ground state one-particle density matrices and the 
corresponding density profiles of the atoms. The density profile of the component $\sigma$ is the 
diagonal part of the one-particle density matrix, $n^{(\sigma)}(x) = \rho^{(\sigma)}(x, x)$, with 
$\rho^{(\sigma)}(x, x^\prime) = \langle \eta | \psi_\sigma^\dag(x) \psi_\sigma(x^\prime) | \eta \rangle$ 
for the system in the state $| \eta \rangle$. 

\begin{figure}[!h]
\includegraphics[width=.99\columnwidth]{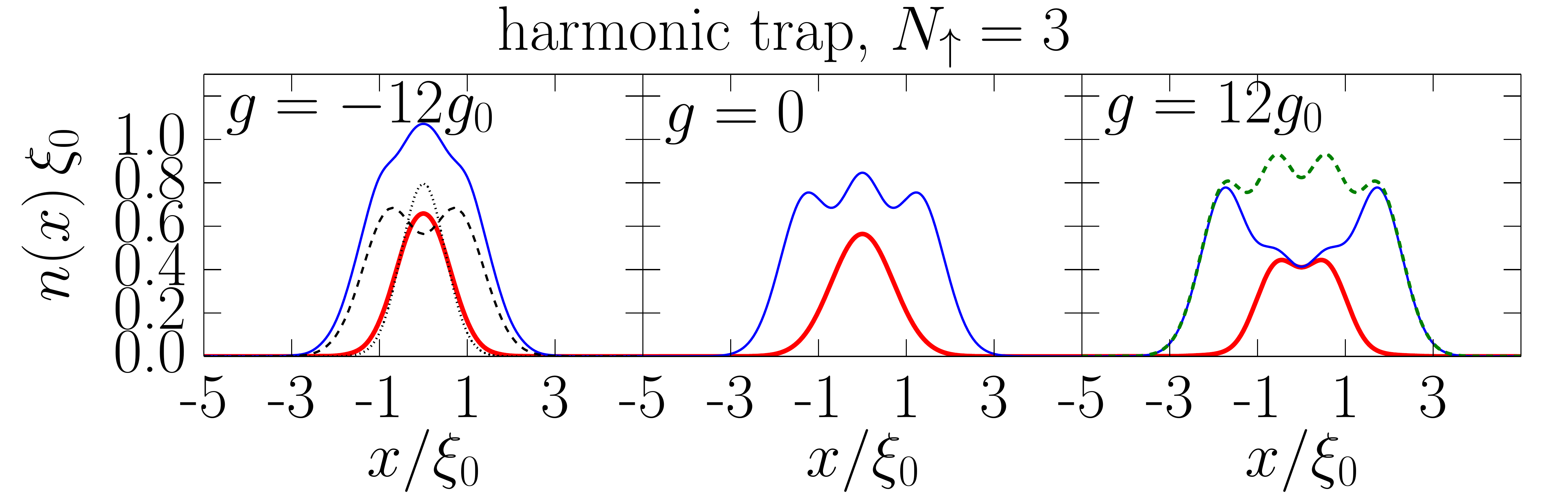}\\
\includegraphics[width=.99\columnwidth]{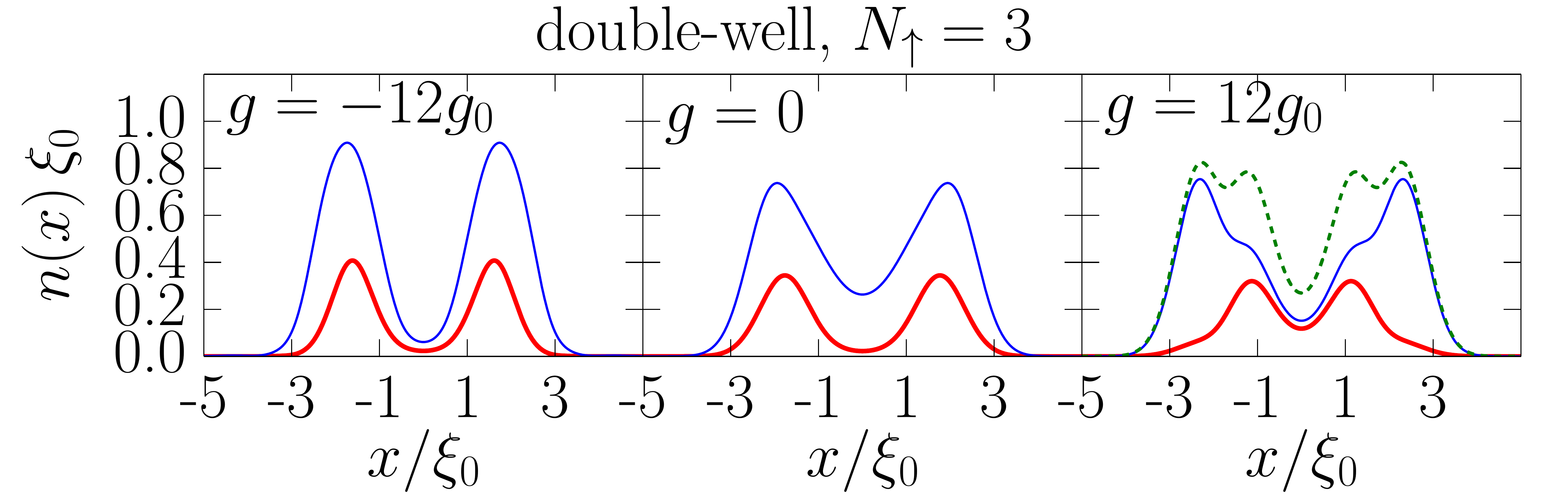}\\
\caption{(color online) Example of densities of three fermions (thin blue line) interacting with an impurity (thick red line). It can be clearly seen, that, for strong repulsion, initially noninteracting impurity starts behaving like the fourth fermion. The total density (shown with a green dashed line), i.e. $n_\uparrow + n_\downarrow$, coincides with that of $N_\uparrow = 4$ non-interacting Fermi gas. For the harmonically trapped system in the strong attraction limit, the density of $N_\uparrow = 2$ Fermi gas and that of a non-interacting molecule (of mass $m_M = 2 m$) are shown (dashed and dotted black lines, respectively). For definitions of the units, see Eq.~(\ref{eq.units}). }
\label{fig.densN3}
\end{figure}

Figure~\ref{fig.densN3} shows the densities of the impurity embedded in three spin-$\uparrow$ fermions for three different regimes of the interaction parameter. In a harmonic trap, when the system becomes non-interacting ($g_{1D} = 0$), the impurity takes the Gaussian shape of a free particle, while the Fermi gas shows three clear peaks of Friedel oscillations corresponding to three spin-$\uparrow$ fermions. For strong attraction, $g_{1D} \to -\infty$, the impurity forms a strong bound state with one fermion. In a homogeneous system it was shown by McGuire in Ref.~\cite{McGuire1966} that the energy in this limit is consistent with that of a molecule which does not interact with $N_\uparrow - 1$ ideal Fermions. Thus the impurity, ``decreases'' the number of ideal fermions by one, as one of the fermions effectively becomes ``distinguishable'' from the others due to the strong binding. Formally, expanding the McGuire expression for the excess energy up to second order in $a$ we get
\begin{equation}
\Delta E \approx -\frac{\hbar^2}{m a^2} - E_F + \frac43 E_F n a + \mathcal{O}(a^2) ~.
\label{eq.McGuireExpansion}
\end{equation}
The first term corresponds to the binding energy of the molecule, while the second terms corresponds to the removal of the bound fermion from the bath. The next-order positive correction can be interpreted as a repulsive interaction between the molecule and the majority fermions, which results in a broader density profile of the two components, as we indeed report in Fig.~\ref{fig.densN3}. In a trapped system the density of a molecule, located at the centre of the trap, is superimposed with that of an ideal Fermi gas of $N_\uparrow - 1$ particles showing the Friedel oscillations, $n_\uparrow(x) = (1 + 2 x^2)\, e^{-x^2} / \sqrt{\pi}$ (dashed black line in Fig.~\ref{fig.densN3}). The molecule, having a mass twice as large as an atom, is expected to be localized at the centre of the trap, enhancing the central peak in the total density, $n_M(x) / 2 = \sqrt{2 / \pi}\, e^{-2 x^2}$ (black dotted line in Fig.~\ref{fig.densN3}). This peak at the trap's centre masks the Friedel oscillation minimum in the density of $N_\uparrow - 1 = 2$ polarized fermions. The larger widths of the profiles can be attributed to the repulsion between the molecule having a double effective mass and effectively reduced in number $\uparrow$-fermions. 

\begin{figure}[!h]
\includegraphics[width=.99\columnwidth]{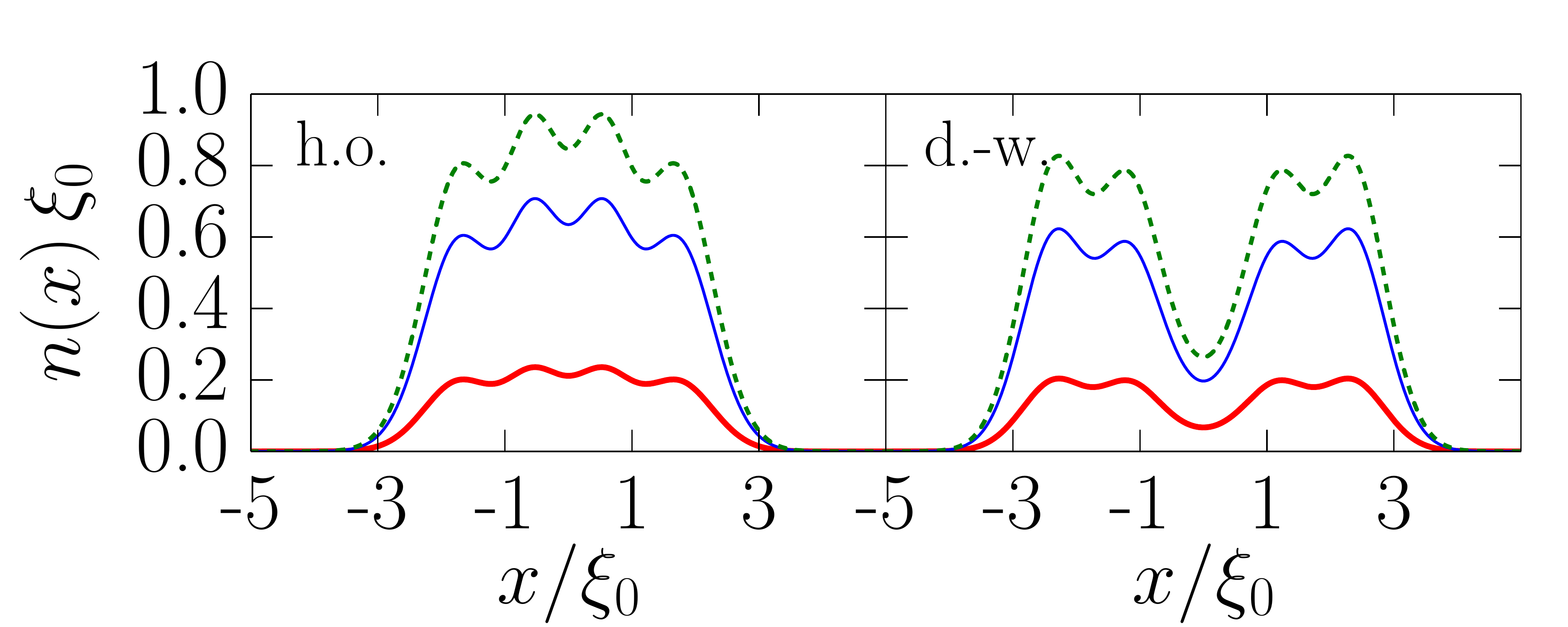}\\
\caption{(color online) Densities of the $N_\uparrow = 3$ antisymmetric state in the degeneracy manifold at $g \to +\infty$ limit for a harmonic trap (left) and a double well (right). The thick (red) line corresponds to the density of the impurity, the thin solid (blue) line to the density of the majority fermions and the dashed line (green) line to the total density $n_\uparrow + n_\downarrow$. }
\label{fig.densAntisymm}
\end{figure}

For strong repulsion, $g_{1D} \to +\infty$, McGuire has shown that in a homogeneous system the energy is comparable to that of $N+1$ ideal fermions \cite{McGuire1965}. Also in a trap a similar effect is observed for the energy, which in this limit becomes degenerate (see Fig.~\ref{fig.spectra}), so that there are states in the degeneracy manifold with different density profiles and the same energy. At the same time, the total density of the entire system now features four peaks instead of three (compare with Refs.~\cite{Gharashi2013,Lindgren2014}) so that the impurity effectively plays a role of an additional ideal fermion. The state which is adiabatically connected to the ground state in $g_{1D}\to-\infty$ limit, where the molecule is localized in the centre, has similarly the impurity localized in the centre. Since in our system the masses are equal, $m_\uparrow = m_\downarrow$, and the $SU(2)$ symmetry is therefore preserved, the Lieb-Mattis theorem holds~\cite{LiebMattis1962}. However, should the mass ratio be different from one, $m_\uparrow / m_\downarrow \neq 1$, the ground state could be interpreted as a few-body counterpart of a ferromagnetic phase separation~\cite{Pecak2016}. 
At the same time the different degenerate states have different symmetries. A fully antisymmetric state, which corresponds to a horizontal line in the spectrum (compare with Fig.~\ref{fig.spectra}), has a density profile corresponding to $N_\uparrow + 1$ non-interacting fermions for both impurity and a majority component, see Fig.~\ref{fig.densAntisymm}. For a finite value of $g>0$, the degeneracy is lifted and the energy is bounded from above by the fully antisymmetric state and from below by the ``molecular'' state in which in the impurity is localized in the centre.

Some of the features are also preserved in the double-well potential. Here, for a high barrier, all particles are localized in the two wells. For strong attraction, $g_{1D} \to -\infty$, this localization is even more prominent than in the non-interacting case, with very little tunnelling between the wells. The difference from the harmonic oscillator case is that the molecule can no longer stay in the centre of the trap, resulting in a degeneracy between the impurity staying in the left and the right well. For strong repulsion, $g_{1D} \to +\infty$, on the other hand, the particles approach the fermionization limit, and the densities add up to a total density of $N_\uparrow + 1 = 4$ non-interacting fermions seen as two Friedel oscillations of the density in each well. The ground state spin-resolved densities remain, however, different for the two components, as was the case for the harmonic trap. Similarly to the harmonic oscillator case, there is a degeneracy between different states with their energy bound from above by the constant energy of the fully anitsymmetric state.

\begin{figure}[!h]
\includegraphics[width=.99\columnwidth]{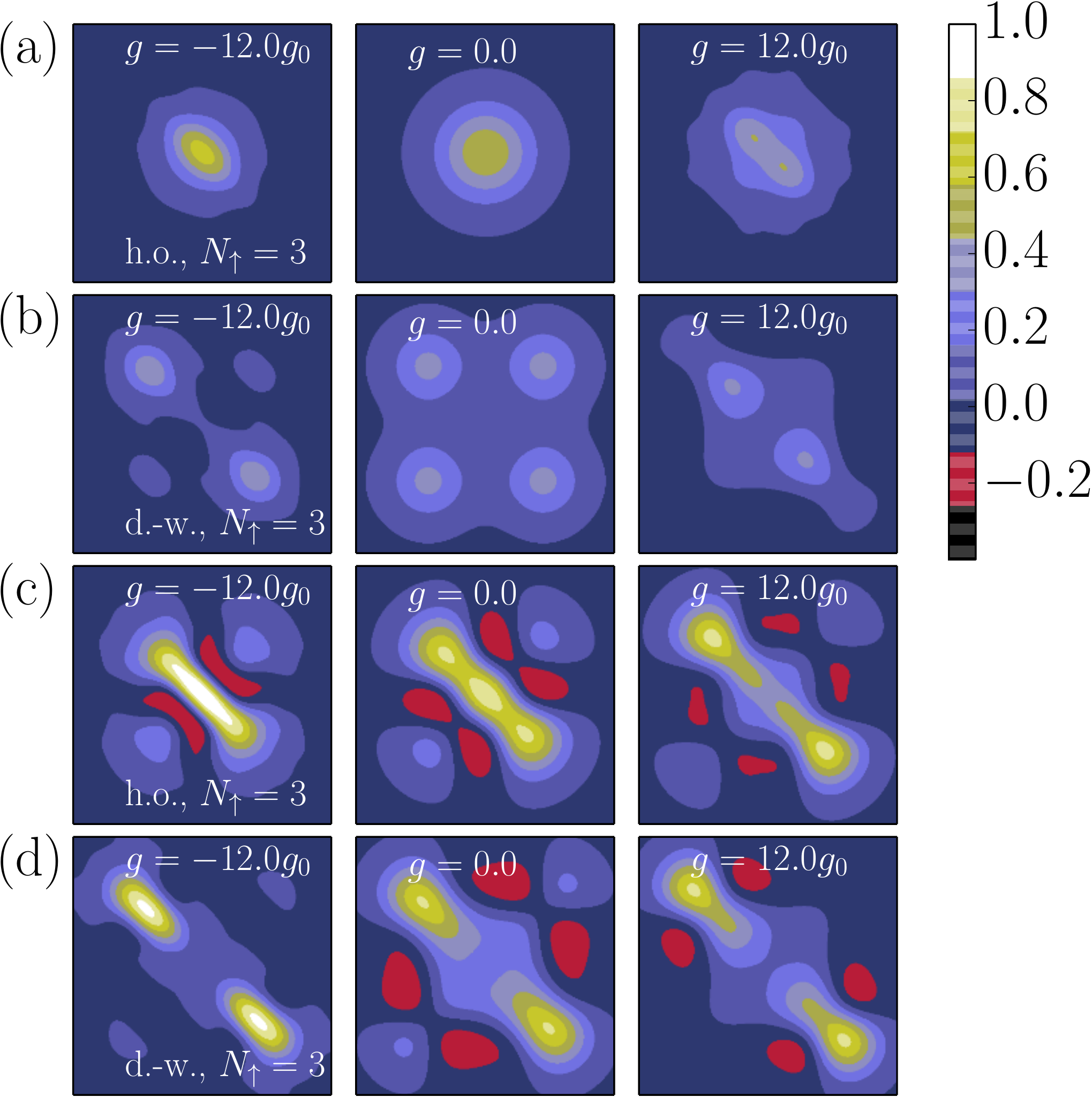}\\
\caption{(color online) Off-diagonal correlation matrices of the polaron, $\rho^{(\downarrow)}(x, x^\prime)$, with $N_\uparrow = 3$ background polarized fermions. Row (a) shows a harmonically trapped system and row (b) a double-well confinement. Rows (c) and (d) show the correlation matrices of the polarized gas, $\rho^{(\uparrow)}(x, x^\prime)$, for a harmonic trap and double-well potential respectively. In the strong repulsion case we can see distinct peaks along a diagonal corresponding to the Friedel oscillations of the density. Higher intensities (brighter colours) correspond to higher values of the one-particle density matrix. In each panel, axis ranges for both $x$ and $x^\prime$ vary from $-4 \xi_0$ to $4 \xi_0$. }
\label{fig.correlfull}
\end{figure}

\begin{figure}[!h]
\includegraphics[width=.99\columnwidth]{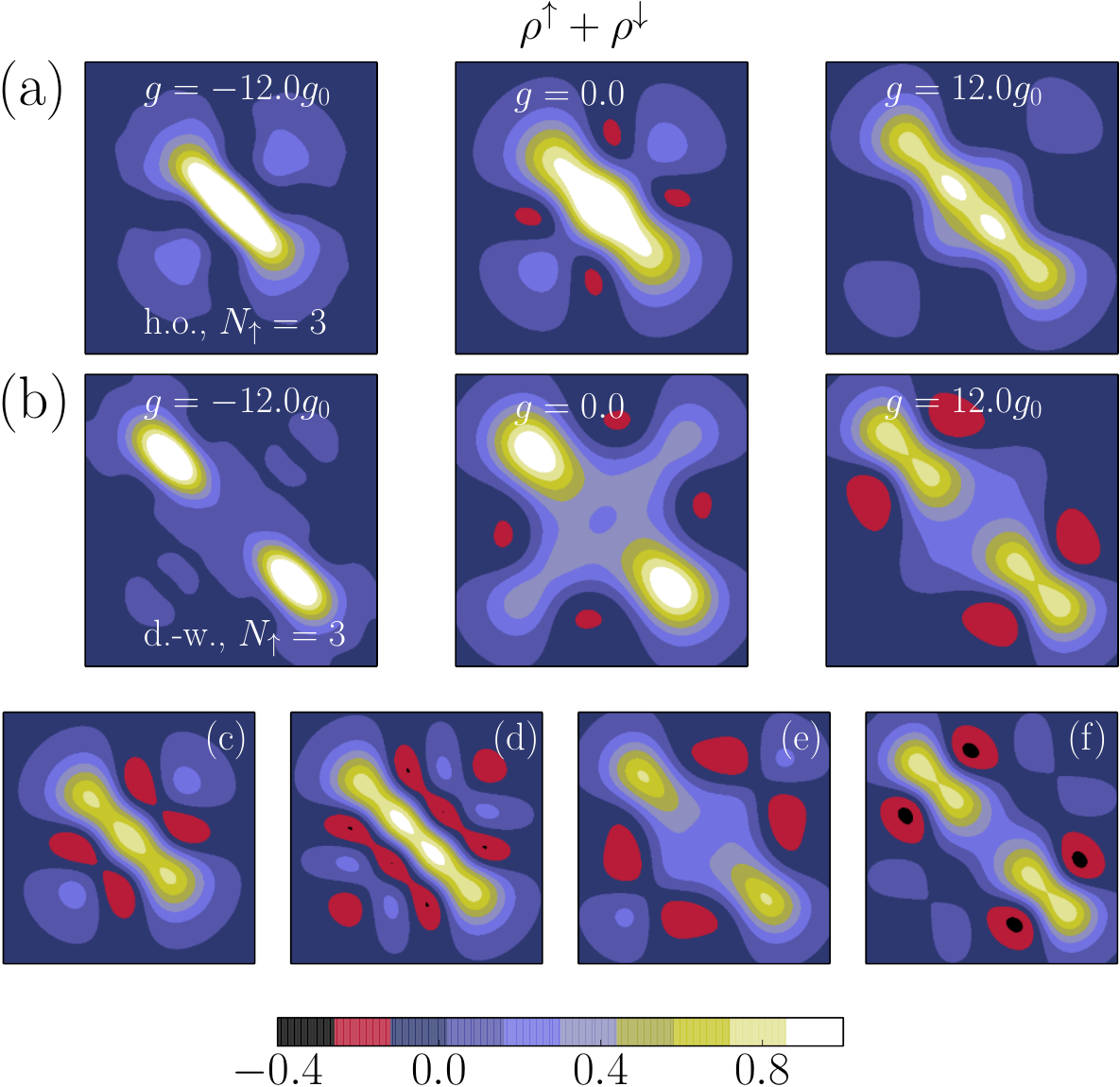}\\
\caption{(color online) Total one-particle density matrices, $\rho^{(\uparrow)}(x, x^\prime) + \rho^{(\downarrow)}(x, x^\prime)$ for a harmonic trap (a) and a double-well potential (b); the correlation matrices of a non-interacting Fermi gas are shown below; (c) for a harmonic trap with $N_\uparrow = 3$, (d) harmonic trap with $N_\uparrow = 4$, (e) double-well with $N_\uparrow = 3$ and (f) double-well with $N_\uparrow = 4$. The off-diagonal elements reveal the difference in behaviour of the strongly repulsive impurity interacting with $N_\uparrow = 3$ polarized Fermions and $N_\uparrow = 4$ non-interacting Fermi gas. Axis ranges the same as in Fig.~\ref{fig.correlfull}. }
\label{fig.correlpg}
\end{figure}

Further understanding of the strongly interacting regime is obtained by inspecting the full one-particle reduced density matrices. Since the Hamiltonian~(\ref{HED}) conserves the number of particles for each spin separately, and therefore $\langle \psi_\downarrow^\dag \psi_\uparrow \rangle = 0$, the total density matrix of the system  equals $\rho^{(\uparrow)}(x, x^\prime) + \rho^{(\downarrow)}(x, x^\prime)$. In the off-diagonal terms of one-particle density matrices $\rho^{(\uparrow)}(x, x^\prime)$ and $\rho^{(\downarrow)}(x, x^\prime)$ reported in Fig.~\ref{fig.correlfull} and in $\rho^{(\uparrow)}(x, x^\prime) + \rho^{(\downarrow)}(x, x^\prime)$ reported in Fig.~\ref{fig.correlpg} one can clearly see the difference between the fermionized impurity and a system of $N_\uparrow + 1$ non-interacting fermions. 

In the case of $g_{1D} = 0$ the correlation function of the impurity can be simply calculated as $\rho^{(\downarrow)}(x, x^\prime) = \varphi^*_0(x) \varphi_0(x^\prime)$ and of the polarized Fermi gas as $\rho^{(\uparrow)}(x, x^\prime) = \sum_{n = 0}^{N_\uparrow - 1} \varphi^*_n(x) \varphi_n(x^\prime)$. We plot it for comparison with the total one-particle density matrix of the interacting system in Fig.~\ref{fig.correlpg}(c)-(f).
It is clear that the structure of the off-diagonal elements of total one-particle density matrix, $\rho^{(\uparrow)}(x, x^\prime) + \rho^{(\downarrow)}(x, x^\prime)$, is different from that of $N_\uparrow + 1$ non-interacting fermions. These are signatures of the distinguishability between the strongly repulsive impurity and the majority fermions.

\section{Dynamics}
\label{sec.oscill}
\subsection{Oscillations in time}
\begin{figure}[!h]
\includegraphics[width=.99\columnwidth]{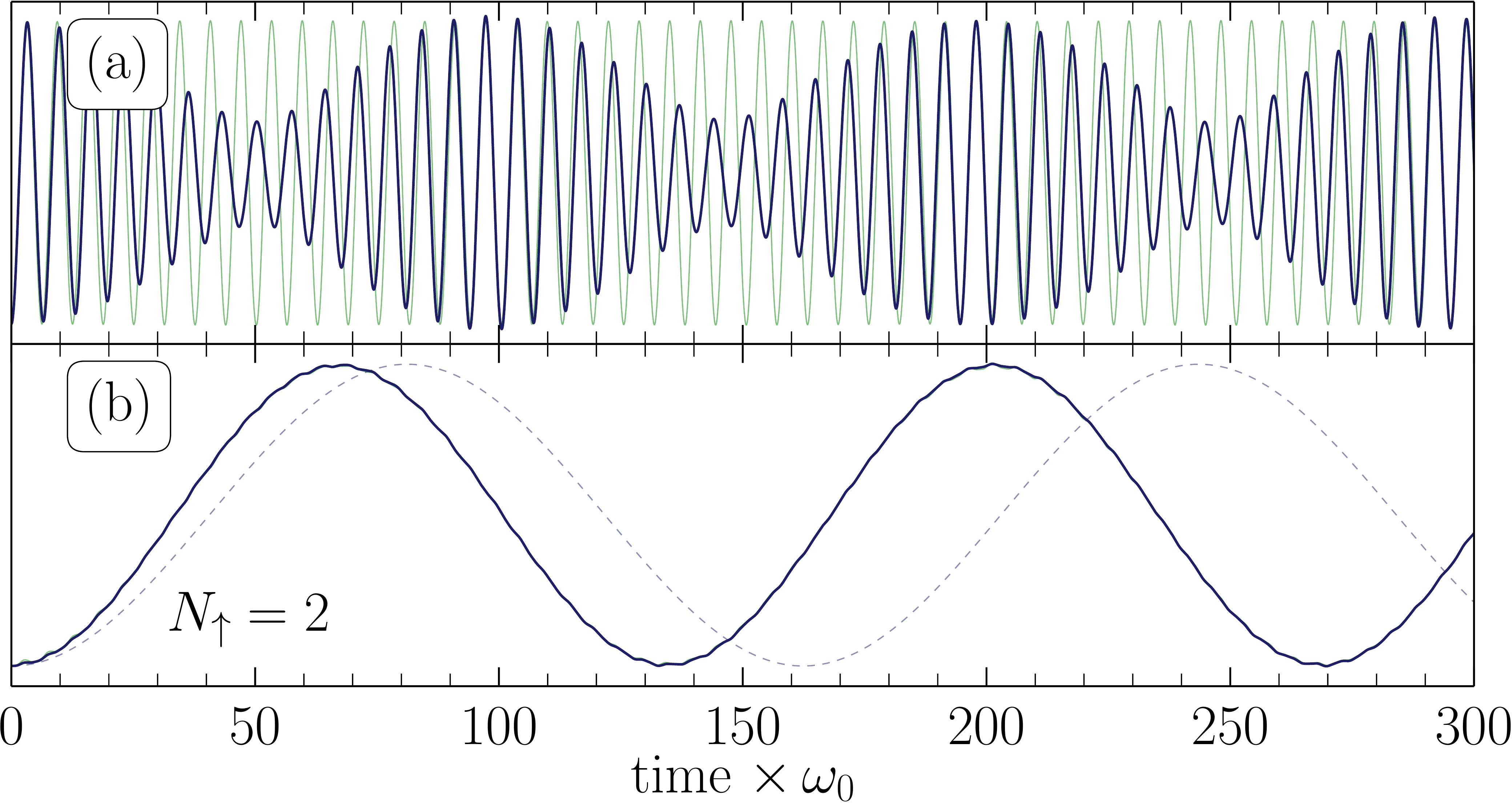}
\caption{(color online) The figure shows oscillations of $\langle x_{\downarrow}(t) \rangle$ in time. In (a) we present the oscillations in a harmonic trap. The faint green line shows the sum of $\langle x_{\downarrow}(t) \rangle + \sum_i \langle x_i(t) \rangle$, i.e. the centre-of-mass motion. As expected, it adds up to a perfect harmonic oscillation with a trap frequency $\omega_0$. In (b) the oscillations in a double-well potential, with much lower frequency due to a barrier tunnelling ($V_0 = 5 \hbar \omega_0$). In both cases the solid line corresponds to the interaction $g_{1D} = 0.4\, g_0$. In panel (b) the dashed line represents $g_{1D} = 0$ for comparison. }
\label{fig.tdep}
\end{figure}

Since our system is small, it allows for a thorough numerical investigation of its out-of-equilibrium properties; in particular the tunnelling dynamics of the polaron. To this end, we initialize the system as a product state with the impurity localized mostly on one side of the trap (in one of the wells in the case of the double well), i.e. the impurity as a linear combination of the two lowest lying single-particle states, $( | 0 \rangle_{\downarrow} - | 1 \rangle_{\downarrow} ) / \sqrt{2}$; the majority atoms are in their non-interacting ground state,
\begin{equation}
| {\text{init}} \rangle = \frac{1}{\sqrt2} \big( | 0 \rangle_{\downarrow} - | 1 \rangle_{\downarrow} \big)\, \otimes| {\rm ground\, state} \rangle_{\uparrow} ~.
\label{eq.init}
\end{equation}
The system is then let evolve with the full Hamiltonian Eq.~(\ref{HED}). For completeness and for comparison we also study the same quench but for a polaron oscillating in a harmonic trap. Figure~\ref{fig.tdep}(a) shows the mean position of the impurity, $\langle x_{\rm \downarrow}(t) \rangle$, as it evolves in time in a harmonic trap. The beats indicate the interference of two frequencies. Indeed, the impurity mode would be a combination of the centre-of-mass motion with frequency $\omega_0$ and of the relative motion with frequency $\omega_1$. These two frequencies are responsible for beats observed in the time evolution of $\langle x_{\rm \downarrow}(t) \rangle$, with frequencies of $(\omega_1 + \omega_0) / 2$ and $|\omega_1 - \omega_0| / 2$.  It is easy to extract $\omega_1$ carrying information about the bath-impurity correlations. As a check, the centre-of-mass position $\sum_{i = 1}^{N_\uparrow} \langle x_i \rangle + \langle x_{\rm \downarrow} \rangle$, shown with a faint line in Fig.~\ref{fig.tdep}(a), corresponds to a harmonic oscillation with $\omega = \omega_0$, i.e. the Kohn mode~\cite{Kohn1961}.
In Fig.~\ref{fig.tdep}(b) we present the evolution of mean impurity position, $\langle x_{\rm \downarrow}(t) \rangle$, in the double-well trap with the same coupling strength as in panel (a), $g_{1D} = 0.4 g_0$ (the dashed line shows the oscillations for $g_{1D} = 0$). In this case, the impurity oscillations are due to tunnelling through a barrier, therefore they are much slower and without beats.

\begin{figure}[!h]
\includegraphics[width=.99\columnwidth]{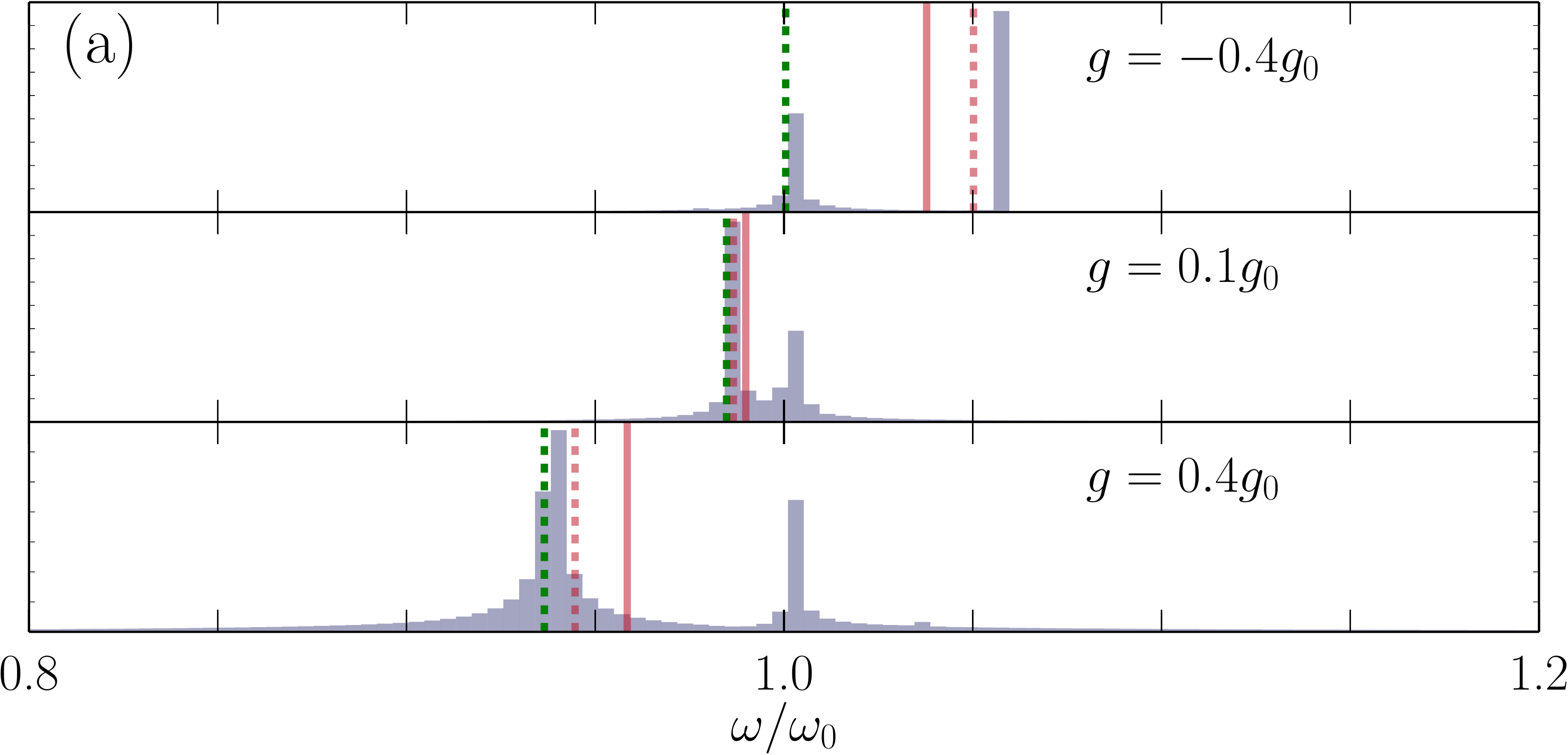}\\
\includegraphics[width=.99\columnwidth]{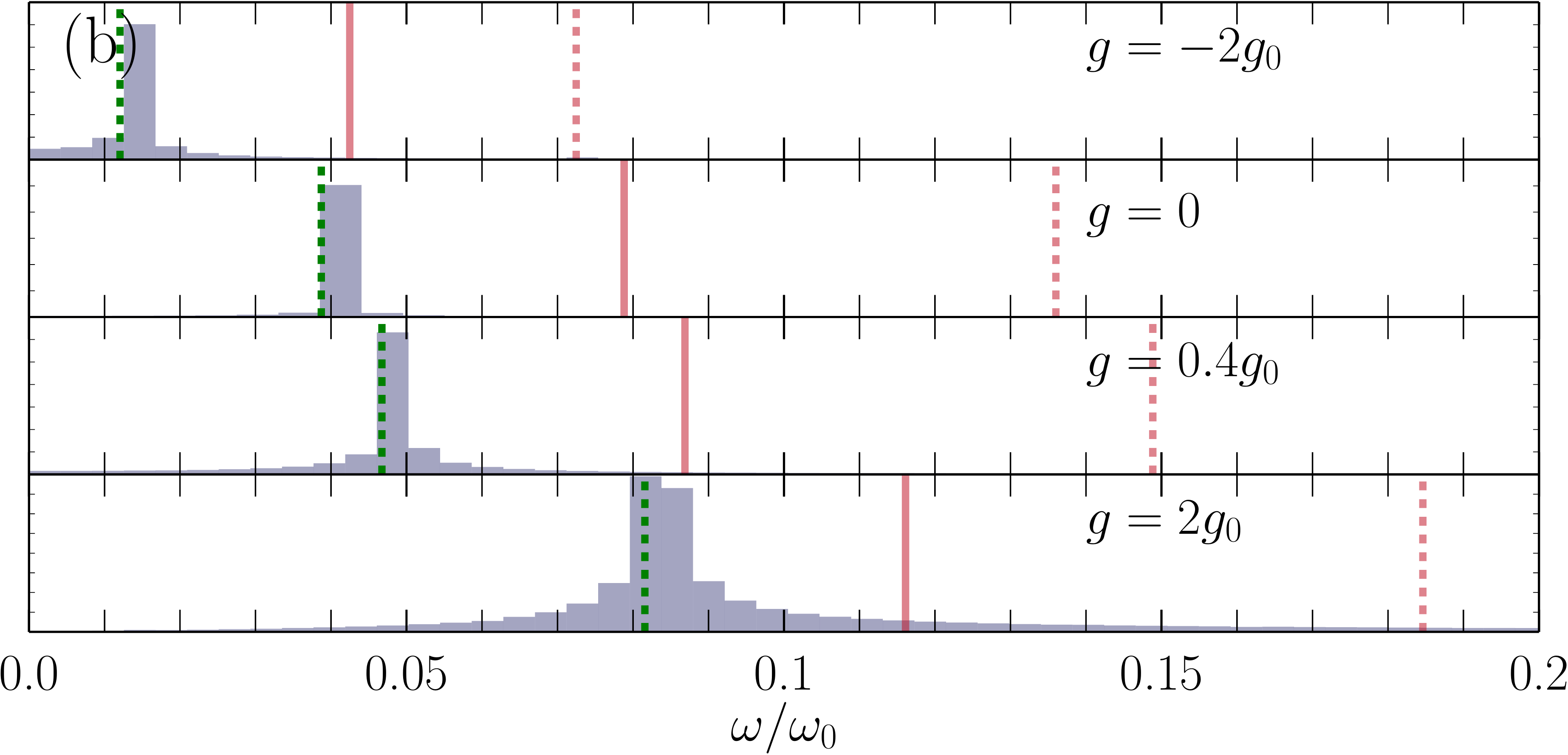}\\
\caption{(color online) The Fourier transform of the time evolution of the state Eq.~(\ref{eq.init}) of $N_\uparrow = 2$ for both (a) the harmonic trap and (b) the double-well potential with $V_0 = 5 \hbar \omega_0$. The peaks seem broader then in panel (a) because of a different scale of the frequency range plotted. The vertical green dashed lines show the oscillation frequencies $\delta \epsilon$ calculated from the exact diagonalization  according to Eq. (\ref{eq.omega1}) in panel
(a) , and Eq. (\ref{eq.omega2}) in panel (b). The red lines show the frequency estimate $\Omega_{-1}$, see Eq.~(\ref{eq.sumrules}), for $\mathcal{F} = X_\downarrow$ (solid lines) and $X_\downarrow - X_\uparrow$ (dashed lines) respectively. }
\label{fig.fft}
\end{figure}

In order to get further insight into the frequency spectrum, we perform a Fourier transform of the time series, $\tilde{x}_\downarrow(\omega) = \int dt\, e^{-i\omega t} \langle x_{\downarrow}(t) \rangle$, and extract the oscillation frequencies. We start with the harmonic potential. The frequency spectrum in this case is shown in Fig.~\ref{fig.fft}(a) and contains two aforementioned frequencies corresponding to the oscillation of the centre-of-mass and to the relative motion of the particles. For attractive (repulsive) interactions, $\omega_1$ is larger (smaller) than $\omega_0$, since it requires  more (less) energy to separate the two species of particles, therefore increasing (decreasing) the restoring force for the relative motion. It is worth noticing that the signal for the out-of-phase mode is very sharp for the attractive interaction, while it is much broader for the repulsive one.

In the double-well trap, see Fig.~\ref{fig.fft}(b), the centre-of-mass motion does not perform harmonic oscillations, and the system has a single oscillation frequency.  Again, the behaviour with respect to the non-interacting case (Fig. \ref{fig.fft}(b) for $g=0$) is opposite for attractive and repulsive interactions. In this case, however, attraction (repulsion) corresponds to a lower (higher) frequency, since the impurity prefers to localize  more (less) in the single well, decreasing (increasing) the splitting between the two lowest energy levels. Finally, if the interaction is very large, the oscillations lose their regularity, and the corresponding peak in the Fourier transform significantly broadens.

The splitting between the low-lying energy levels of the spectrum,
\begin{subequations}
\begin{eqnarray}
\label{eq.omega1}
\hbar \omega &=& \epsilon_1 - \epsilon_0 ~,\\
\hbar \omega &=& \epsilon_2 - \epsilon_1 ~,
\label{eq.omega2}
\end{eqnarray}
\end{subequations}
can give a good estimate of the oscillation frequency of the impurity, $\omega$. Physically, it is similar to the oscillation of a two-level system, whose time evolution is given by $e^{i \epsilon_0 t} (| 0 \rangle - e^{i(\epsilon_1 - \epsilon_0)t} | 1 \rangle) / \sqrt2$. Depending on the symmetry of oscillations (centre-of-mass or relative motion), splitting between relevant energy levels should be considered. In Fig. \ref{fig.fft} we show that this level splitting coincides with the lowest frequency of the Fourier spectrum (vertical green dashed lines).

\subsection{Sum rules approach}
\label{subsec.sumrules}
A powerful method to estimate and understand the frequencies of different modes of a system is the so called sum-rule approach (see, e.g., ~\cite{BlackBook2016} and reference therein for its application to ultracold gases). The method is based on determining the frequency moments of the structure factor
\begin{equation}
S_{\mathcal{F}}(E) = \sum_\eta\, |\langle \eta | \mathcal{F} | \eta_0 \rangle|^2\, \delta(E + \epsilon_0 - \epsilon_\eta),
\label{eq.sf}
\end{equation}
for a certain operator $\mathcal{F}$ corresponding to the mode of interest. In Eq.~(\ref{eq.sf}) $E$ is the energy of the excitation, $| \eta \rangle$ are the eigenstates of the full Hamiltonian in Eq.~(\ref{eq.hamilt}), $\mathcal{H} | \eta \rangle = \epsilon_\eta | \eta \rangle$, while $| \eta_0 \rangle$ denotes its ground state. Since we are interested in the position of the atoms, we consider as $\mathcal{F}$ the operators of position of the impurity and of the polarized gas,
\begin{equation}
X_\sigma = \int_{-\infty}^\infty dx\, x\, \psi_\sigma^\dag(x) \psi_\sigma(x) ~,
\end{equation}
or their combinations as a dipole and a centre-of-mass operators.
\begin{equation}
F_\pm = X_\downarrow \pm X_\uparrow ~.
\end{equation}

Within the sum-rule approach the oscillation frequency can be estimated from the ratio of the moments as
\begin{equation}
\hbar \Omega_{-1} = \sqrt{m_1 / m_{-1}} ~,
\label{eq.sumrules}
\end{equation}
where the moments of the structure factor are defined as
\begin{equation}
m_k(\mathcal{F}) = \int_{-\infty}^{\infty} S_\mathcal{F}(E) E^k\, dE ~.
\end{equation}
For our operators $\mathcal{F}$, $m_1 = \frac12 \langle [\mathcal{F},[H,\mathcal{F}]] \rangle$ does not depend on the interaction and corresponds to the f-sum rule. On the other hand, $m_{-1}$ is proportional to the susceptibility related to the operators $\mathcal{F}$. The behaviour of the collective mode frequency $\omega_1$ in a harmonic trap is easily explained, since $m_{-1}(F_{-})$ is proportional to the spin susceptibility \cite{AleSandroSpin}. For an attractive (repulsive) interaction  the susceptibility decreases (increases) with respect to its non-interacting value. Indeed, in Fig. \ref{fig.fft} the sum rule approach for the operator $F_{-}$ (red dashed line) gives a good account of $\omega_1$, showing that this frequency corresponds to the out-of-phase mode. While the impurity motion (red solid line) takes contribution from both the centre of mass and the relative motion. On the other hand, our operator $F_{-}$ is not well defined in the double well, since the centre-of-mass and the relative motion are coupled. As expected, in this case, the sum rule gives a better result when considering the single impurity operator $X_\downarrow$.

\section{RF spectroscopy}
\label{sec.rabi}
\begin{figure}[!h]
\includegraphics[width=.99\columnwidth]{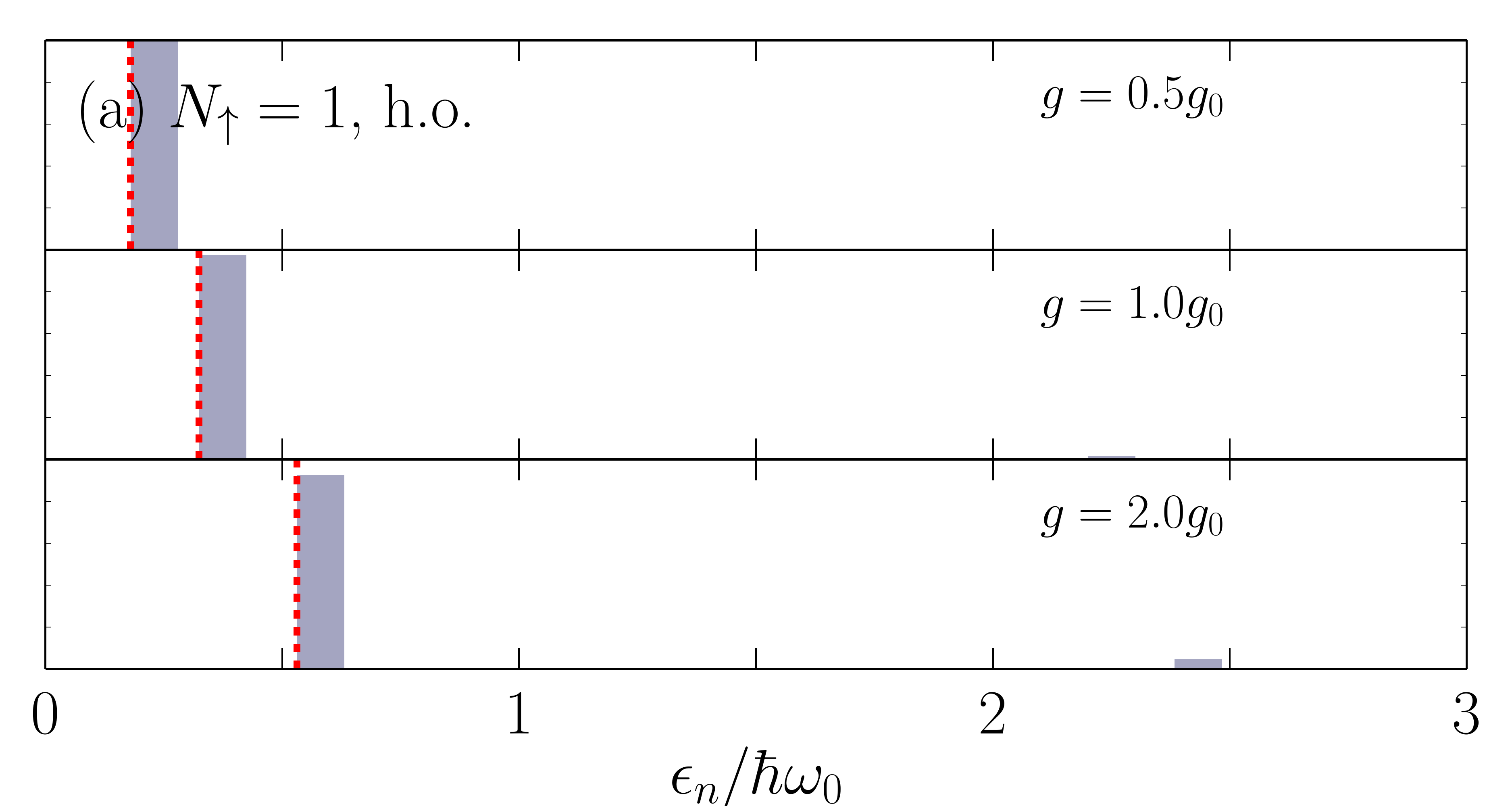}\\
\includegraphics[width=.99\columnwidth]{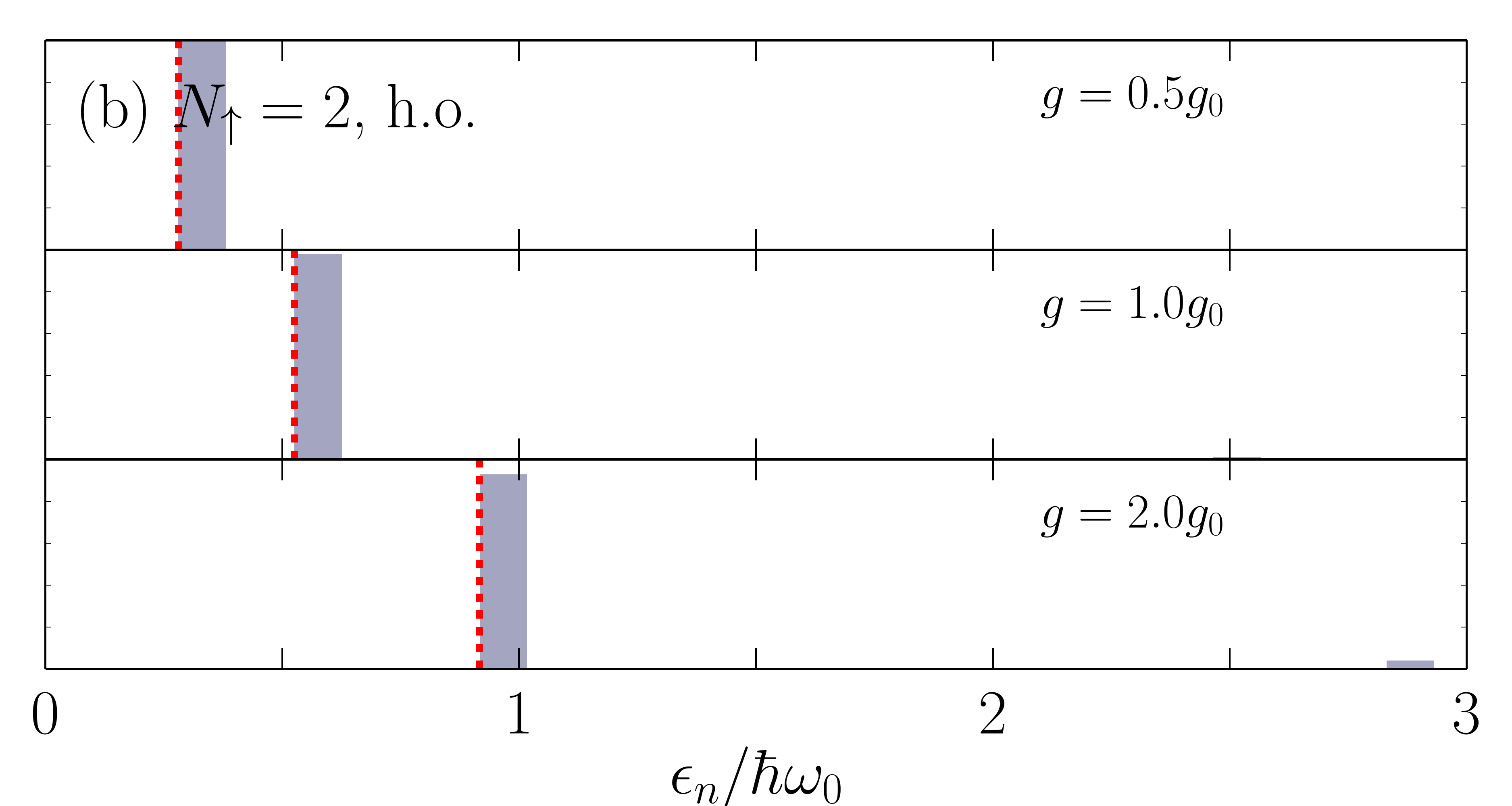}\\
\caption{(color online) The spectral function $S_\Omega$,~(\ref{eq.strfact_rabi}) in a harmonic oscillator potential: (a) for $N_\uparrow = 1$ and (b) for $N_\uparrow = 2$. Only odd peaks are non-zero due to the parity of the system, and get exponentially suppressed for higher energies. The red lines show the corresponding polaron energies calculated with exact diagonalization, see Fig~\ref{fig.deltaEho}.}
\label{fig.strfact_rabi_ho}
\end{figure}
\begin{figure}[!h]
\includegraphics[width=.99\columnwidth]{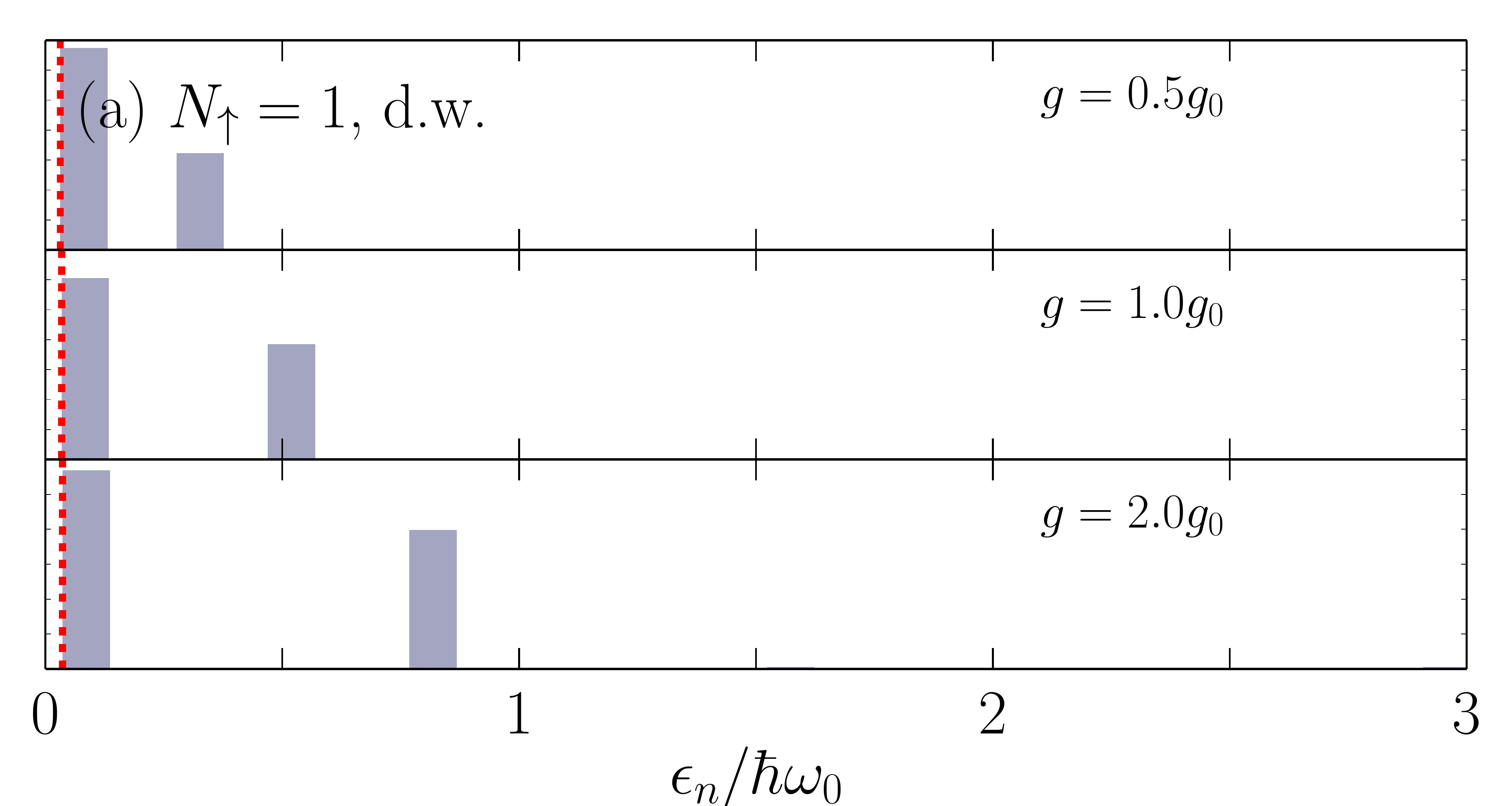}\\
\includegraphics[width=.99\columnwidth]{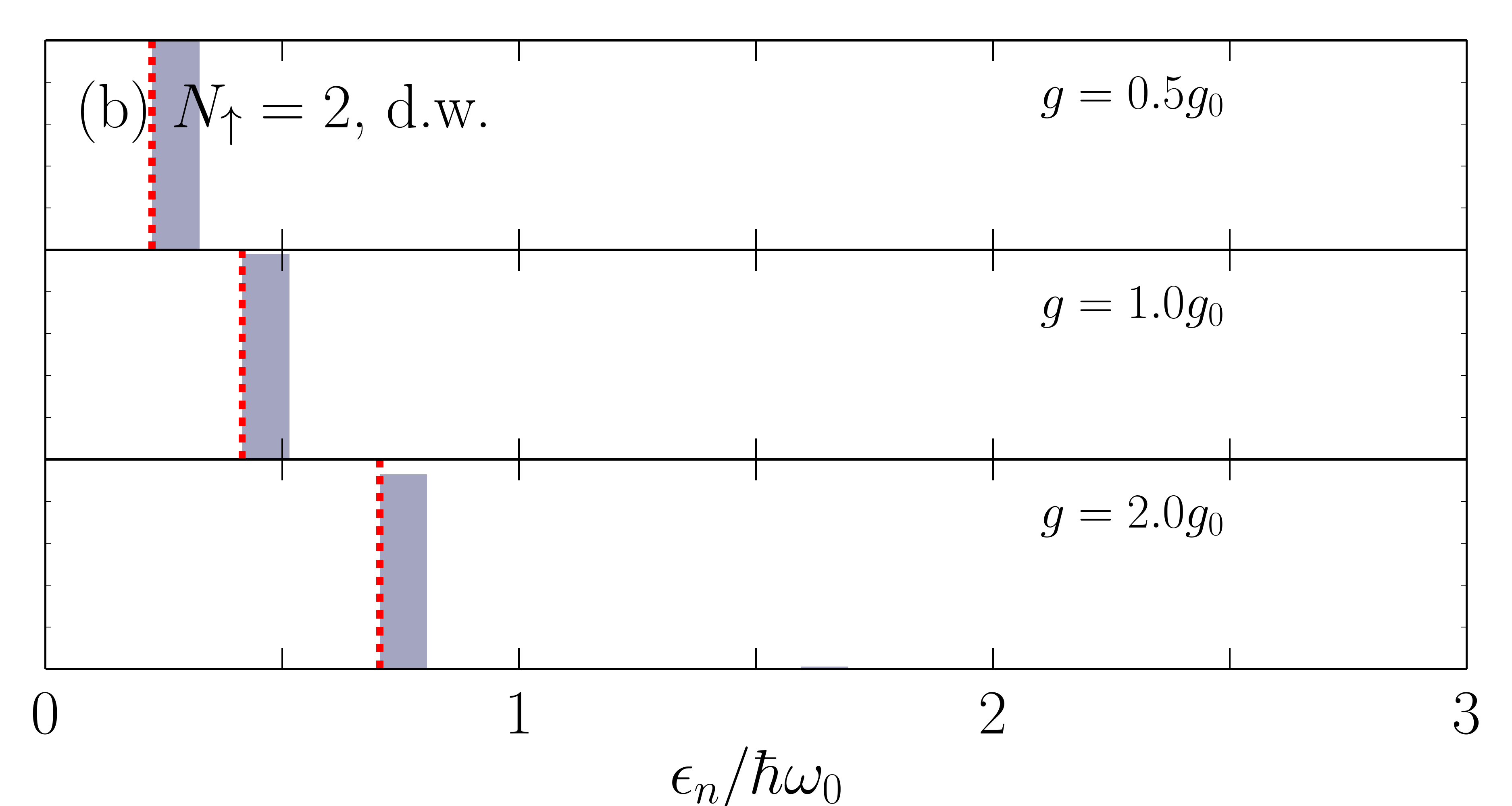}\\
\caption{(color online) The spectral function~(\ref{eq.strfact_rabi}) in a double well potential: (a) for $N_\uparrow = 1$ and (b) for $N_\uparrow = 2$. In the first case there are two peaks, where the first one arises because the two lowest levels are almost degenerate. In the latter case, the second peak is already farther apart and thus exponentially suppressed. The red lines show the corresponding polaron energies calculated with exact diagonalization, see Fig~\ref{fig.deltaEdw}.}
\label{fig.strfact_rabi_dw}
\end{figure}
One of the most powerful methods to characterize the polaron experimentally is the radio-frequency (RF) spectroscopy (see, e.g. \cite{ZwergerVarenna2016} and references therein). With this method, the spectral function $I(E)$ of the impurity is measured by applying an RF pulse to the impurity. This pulse changes its internal (hyperfine) state. When a continuous signal is applied instead of a pulse, it is possible to induce Rabi oscillations of the impurity, which also carry information about the coherence of the polaron, and from which it is possible to extract information similar to that obtained by measuring the spectral function~\cite{RudiTeo,Scazza2016}.

We consider a system where the impurity has two internal states, which we label as $|\downarrow 2\rangle$ and $|\downarrow 3\rangle$, and we add a Rabi coupling term to the Hamiltonian,
\begin{equation}
\mathcal{H}_{\Omega} = \frac12 \Omega_R \int dx\; \Phi^\dag(x) \sigma_x \Phi(x) ~,
\end{equation}
with $\Phi(x) = (\psi_{\downarrow 2}(x), \psi_{\downarrow 3}(x))^T$ and $\sigma_x$ the first Pauli matrix. This term allows for oscillations of the impurity between the state that interacts with the polarized gas and the non-interacting state. Similarly to the case studied in previous paragraphs we may decompose the field operators into single-particle modes $\psi(x)_{\downarrow \alpha} = \sum_n a_{\downarrow \alpha, \, n} \varphi_n(x)$ with $\alpha = 2, 3$, and where $a_{\downarrow 2}$ and $a_{\downarrow 3}$ are annihilation operators of the two internal states of the impurity. In this basis, the new Hamiltonian reads
\begin{align} \nonumber
\mathcal{H} = \sum_i \epsilon_i (a_{\uparrow i}^\dag a_{\uparrow i} + a_{2\downarrow i}^\dag a_{2\downarrow i} +  a_{3\downarrow i}^\dag a_{3\downarrow i})\\ \nonumber
+ \sum_{ijkl} J_{ijkl}\, a_{\uparrow i}^\dag a_{3\downarrow j}^\dag a_{3\downarrow k} a_{\uparrow l}\\
+ \frac12 \Omega_R \sum_i ( a_{2\downarrow i}^\dag a_{3\downarrow i} + a_{3\downarrow i}^\dag a_{2\downarrow i} ) ~.
\label{eq.rabi}
\end{align}
We assumed as an approximation that only one of the states of the impurity, labelled here as $|\downarrow 3\rangle$, interacts with the polarized gas.

The spectral function is related to the response of the system to the term $\mathcal{H}_\Omega$ in the Hamiltonian, i.e. to the absorption of the RF beam in the atomic cloud. The latter can be determined by {\it Fermi's Golden Rule}, taking the impurity to be initially in the noninteracting state $| \downarrow 2 \rangle$. Therefore, the response of the system  to $\mathcal{H}_\Omega = \frac12 \Omega_R \sum_i a_{\downarrow 3, i}^\dag a_{\downarrow 2, i} + {\rm H.c.}$ reads
\begin{align}\nonumber
S_{\Omega}(E) \propto \sum_\eta |\langle \eta | \mathcal{H}_\Omega | \eta_0 \rangle|^2 \delta(E - \epsilon_\eta + \epsilon_0) \\
= |\Omega_R|^2 I(E) ~,
\label{eq.strfact_rabi}
\end{align}
where $| \eta \rangle$ are the eigenstates of the Hamiltonian~(\ref{eq.rabi}) with $\Omega_R = 0$ (the interacting eigenstates) and $| \eta_0 \rangle$ is the ground state. The results are shown in Figs.~\ref{fig.strfact_rabi_ho}~and~\ref{fig.strfact_rabi_dw}. At $g_{1D} = 0$, there will be only one peak at zero energy, as the matrix element in Eq.~(\ref{eq.strfact_rabi}) will be proportional to $\delta_{\eta \eta_0}$. For interacting system the spectral function will be, in general, given by a coherent polaron peak and an incoherent particle-hole spectrum. For our small system and for a harmonic confinement, the spectral function is dominated by the coherent peak at the polaron energy reported in Fig.~\ref{fig.deltaEho}. For the double well, a sort of closed {\it ``shell''} effect is present. For $N_\uparrow=1$ (aside from the polaron peak, which is very close to zero and almost independent of the interaction due to the almost degenerate symmetric and antisymmetric state) there appears a second peak corresponding to higher energy excitations. Instead, for $N_\uparrow=2$, the scattering is with the closed $\uparrow$ {\it ``shell''}, and the impurity gives a spectral function completely dominated by the polaron peak (see also Fig.~\ref{fig.deltaEdw}). The predicted spectral function should be accessible experimentally, also for our small system.

\bigskip
{\bf Conclusions}.--- We studied the tunnelling properties of the Fermi polaron and its Rabi oscillations in harmonic and double-well traps. Using LDA and exact diagonalization we obtained polaron energies that agree well with the McGuire formula (\ref{eq.mcguire}) generalized to non-uniform systems. Clear steps in the polaron energy are a signature of a double-well potential and can be observed experimentally. We also showed that the dynamics of tunnelling through a barrier can be inferred from the spectrum of the system and from its structure factor. Finally, we calculated the spectrum and Rabi oscillations of an impurity that has two internal states. There is a possibility of an experimental measurement of these quantities; such an experiment should be feasible given the current progress in the field. This in turn should contribute to our understanding of the physics of Fermi polaron and interactions between fermions of two species.

\bigskip
{\bf Acknowledgements}.--- We are grateful to L.P.~Pitaevskii, A.G.~Volosniev, S.~Jochim, T.~Sowi\'{n}ski and D.~P\c{e}cak for stimulating discussions.
This work was supported by the Provincia Autonoma di Trento.
During the final stage of manuscript preparation MT was also supported by ERC Advanced Grant ``Condensation in Designed Systems'' funding 675013.
G.E.A. acknowledges partial financial support from the MICINN (Spain) Grant No.~FIS2014-56257-C2-1-P.
The Barcelona Supercomputing Center (The Spanish National Supercomputing Center -- Centro Nacional de Supercomputaci\'on) is acknowledged for the provided computational facilities. \\

\end{document}